\documentclass{article}

\usepackage[preprint, nonatbib]{neurips_2024}

\usepackage[utf8]{inputenc} 
\usepackage[T1]{fontenc}    
\usepackage{hyperref}       
\usepackage{url}            
\usepackage{booktabs}       
\usepackage{amsfonts}       
\usepackage{nicefrac}       
\usepackage{microtype}      
\usepackage{xcolor}         

\usepackage{subfiles}
\usepackage{amsmath}
\usepackage{algorithm}
\usepackage{algpseudocode}
\usepackage{graphicx}
\usepackage{subcaption}

\usepackage{amsthm}
\newtheorem{theorem}{Theorem}

\usepackage{authblk}
\usepackage{marvosym}

\title{Approximating mutual information of high- dimensional variables using learned representations}

\author[1, 2, *]{Gokul Gowri}
\author[1, 2]{Xiao-Kang Lun}
\author[2, *, \Cross]{Allon M. Klein}
\author[1, 2, \Cross]{Peng Yin}
\affil[1]{Wyss Institute for Biologically Inspired Engineering}
\affil[2]{Department of Systems Biology, Harvard University}
\affil[*]{corresponding authors: ggowri\texttt{@}g.harvard.edu, allon\_klein\texttt{@}hms.harvard.edu}
\affil[\Cross]{senior authors: A.M.K. and P.Y. co-supervised this work.}

\begin{document}

\maketitle

\begin{abstract}
  Mutual information (MI) is a general measure of statistical dependence with widespread application across the sciences. However, estimating MI between multi-dimensional variables is challenging because the number of samples necessary to converge to an accurate estimate scales unfavorably with dimensionality. In practice, existing techniques can reliably estimate MI in up to tens of dimensions, but fail in higher dimensions, where sufficient sample sizes are infeasible. Here, we explore the idea that underlying low-dimensional structure in high-dimensional data can be exploited to faithfully approximate MI in high-dimensional settings with realistic sample sizes. We develop a method that we call latent MI (LMI) approximation, which applies a nonparametric MI estimator to low-dimensional representations learned by a simple, theoretically-motivated model architecture. Using several benchmarks, we show that unlike existing techniques, LMI can approximate MI well for variables with $> 10^3$ dimensions if their dependence structure has low intrinsic dimensionality. Finally, we showcase LMI on two open problems in biology. First, we approximate MI between protein language model (pLM) representations of interacting proteins, and find that pLMs encode non-trivial information about protein-protein interactions. Second, we quantify cell fate information contained in single-cell RNA-seq (scRNA-seq) measurements of hematopoietic stem cells, and find a sharp transition during neutrophil differentiation when fate information captured by scRNA-seq increases dramatically.

\end{abstract}

\section{Introduction}

Mutual information is a universal dependence measure which has been used to describe relationships between variables in a wide variety of complex systems: developing embryos \cite{Dubuis2013-yw}, artificial neural networks \cite{Shwartz-Ziv2017-us}, flocks of birds \cite{Meijers2021-ml}, and more. Its widespread use can be attributed to at least two of its appealing properties: equitability and interpretability. 

Many dependence measures are inequitable, meaning that they are biased toward relationships of specific forms \cite{Reshef2011-xx}. For example, Pearson correlations quantify the strength of linear relationships, and Spearman correlations quantify the strength of monotonic relationships. Inequitability can be particularly problematic for complex systems, where relationships can be nonlinear, non-monotonic, or involve higher-order interactions between multidimensional variables \cite{Thibeault2024-an}. Mutual information (MI) stands out as an equitable measure that can capture relationships of any form, and generalizes across continuous, discrete, and multidimensional variables \cite{Kinney2014-yd}. And when scaled consistently, MI provides a universal currency in interpretable units -- which can be understood as the number of `bits’ of information shared between variables \cite{Cover2006-hn}. MI can also be interpreted through decomposition into pointwise mutual information (pMI) \cite{Tsai2020-yt, Kong2023-rr}, which attributes dependence to specific pairs of values.

MI can be defined as the Kullback-Leibler divergence, $D_{KL}$, of a joint distribution from the product of its marginals. For absolutely continuous random variables $X, Y$ defined over $\mathcal{X, Y}$, with joint distribution $P_{XY}$ and marginal distributions $P_X, P_Y$

\begin{equation}
I(X;Y) = D_{\text{KL}} (P_{XY} || P_X \otimes P_Y ) = \int_{\mathcal{X}} \int_{\mathcal{Y}} P_{XY} (x, y) \log \frac{P_{XY} (x, y)}{P_{X} (x) P_{Y} (y)} dy dx
\end{equation}

In practice, $P_{XY}$ is often unknown, and $I(X;Y)$ must be estimated from observations $\{(x_i, y_i)\}$ that sparsely sample $P_{XY}$. While nonparametric MI estimators have been remarkably successful for variables with a single dimension \cite{Kraskov2004-sh, Holmes2019-yx, Czyz2023-qs, Chan2017-io}, MI estimation for high-dimensional variables remains a significant challenge. Nonparametric MI estimation suffers from the curse of dimensionality -- accurate estimation requires a number of samples that scales exponentially with the dimensionality of the variables \cite{Goldfeld2021-gz}.

An exciting recent approach to scaling MI estimates to high dimension is the use of variational bounds on KL divergence to reduce the MI estimation problem to a gradient descent optimization problem \cite{Belghazi2018-ik, Poole2019-iu}. MI estimators based on variational bounds indeed empirically perform well for data with ones to tens of dimensions \cite{Czyz2023-qs}, but still suffer from the curse of dimensionality \cite{Goldfeld2021-gz, McAllester2020-vy}, and can exhibit high variance \cite{Poole2019-iu, Song2019-as}. To our knowledge, no techniques have been shown to reliably estimate MI in practice for variables with hundreds or thousands of dimensions -- a regime relevant to many fields, including genomics, neuroscience, ecology, and machine learning \cite{Klein2015-jz, Ganguli2012-vh, Thibeault2024-an, Goldfeld2018-xq}. 

More generally, it has been shown that no technique can accurately estimate MI from finite samples without making strong assumptions about the distribution from which samples are drawn \cite{McAllester2020-vy}, resulting in a fundamental tension between the theoretical appeal of MI and the practical difficulty of its estimation. One way to resolve this tension is to develop alternative measures of statistical dependence, which retain desirable properties of MI, but are feasible to estimate. Sliced MI, which is the average of MI estimates on random low-dimensional linear projections (``slices'') of high-dimensional data, is an example of such an approach \cite{Goldfeld2021-gz, Goldfeld2022-sg}. While sliced MI is an appealing measure for information-theoretic objective functions in machine learning \cite{Tsur2023-li, Chen2023-nj}, it does not retain the interpretability (in bits) of classical MI, and is inequitable, as it quantifies information that can be extracted through linear projections \cite{Goldfeld2021-gz}.

Here, we take a complementary approach to sliced measures. Rather than considering alternatives to classical MI, we ask if it is possible to make strong, yet reasonable, assumptions about data which enable feasible MI estimation. In this work, we explore the usefulness of the empirically supported assumption that complex systems have underlying low-dimensional structure \cite{Thibeault2024-an}.

Specifically, we propose latent mutual information (LMI) approximation, which applies a nonparametric MI estimator to mutually informative compressed representations of high-dimensional variables. To learn such representations, we design a simple neural network architecture motivated by information-theoretic principles. We demonstrate, using synthetic multivariate Gaussian data, that LMI approximation can be stable for variables reaching thousands of dimensions, provided their dependence has low-dimensional structure. We then introduce an approach for resampling real data to generate benchmark datasets of two high-dimensional variables where ground truth mutual information is known. Using this approach, we evaluate the ability of LMI to capture statistical dependence in two types of real data: images and protein sequence embeddings. Finally, we apply LMI to two open problems in biology: quantifying interaction information in protein language model embeddings, and quantifying cell fate information in the gene expression of mouse hematopoietic stem cells.

\section{Approach}

\begin{figure}
    \centering
    \includegraphics[width=0.75\textwidth]{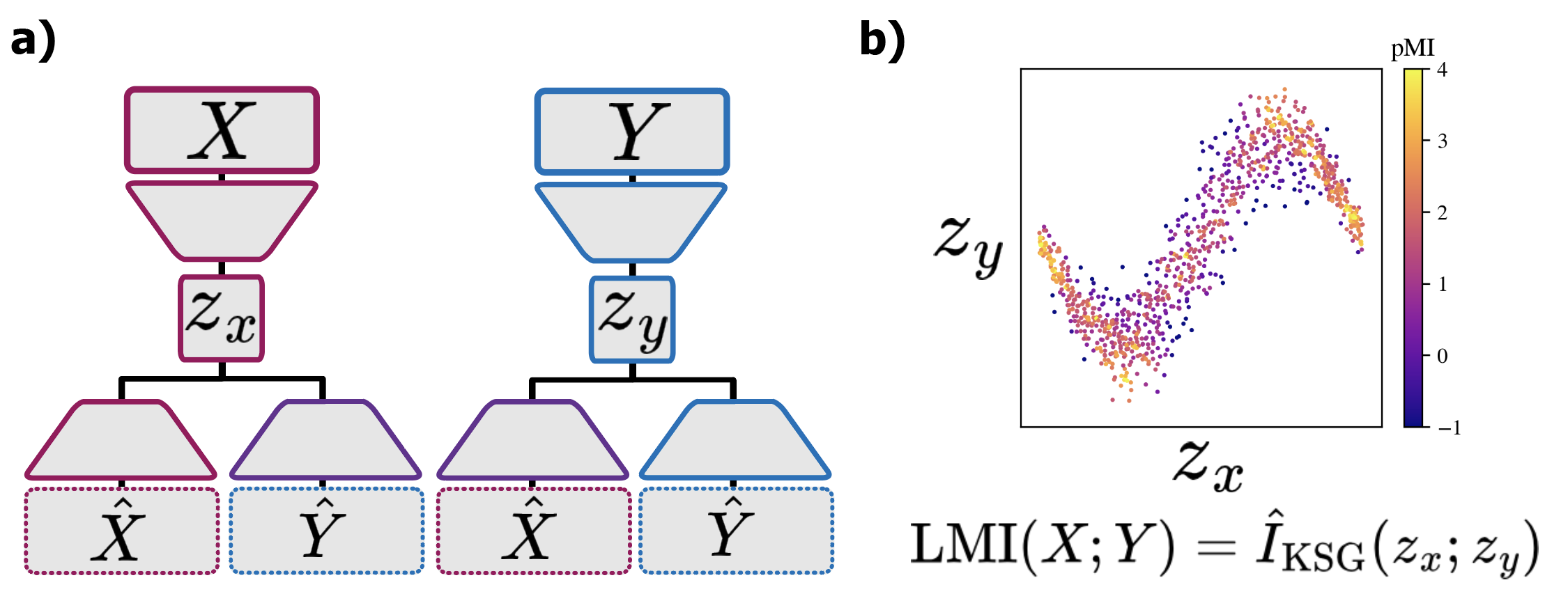}
    \caption{\textbf{Workflow of latent MI approximation} \textbf{a)} Embed high-dimensional data in low-dimensional space such that mutually informative structure is preserved. \textbf{b)} The KSG estimator \cite{Kraskov2004-sh} is used to estimate MI by averaging over pointwise MI (pMI) contributions.}
    \label{fig:enter-label}
\end{figure}


Our goal is to approximate MI from high-dimensional data using low-dimensional representations which capture dependence structure. Our specific approach is to use neural networks to map variables $X, Y$ to low-dimensional representations $Z_x, Z_y$. Then, we use the well-established nonparametric MI estimator introduced in \cite{Kraskov2004-sh} to estimate $\hat{I}(Z_x; Z_y)$.

The central challenge here is to learn $Z_x, Z_y$ such that $\hat{I}(Z_x; Z_y) \approx I(X; Y)$. One sensible approach would be to use autoencoders \cite{Hinton2006-tg} or other popular nonlinear dimensionality reduction techniques \cite{Becht2018-pf, Moon2019-pc} to compress each variable separately. While such an approach could yield a good approximation if compression is perfectly lossless, it can result in a poor approximation if compression is lossy. An illustrative example is two variables each with thousands of independent dimensions but a single pair of strongly dependent dimensions -- two separate lossy compressions are unlikely to preserve the rare dependent components.

We can make this intuition precise using properties of entropy and MI \cite{Cover2006-hn}. For simplicity, let us consider an approximation using only one compressed variable, $Z_y = f (Y)$. By rewriting MI in terms of differential entropy, denoted $h$, we see that for absolutely continuous $X, Y$ with finite differential entropy,

\begin{equation}
    I(X; Y) - I(X; Z_y) = h(X) - h(X | Y) - h(X) + h(X|Z_y) = h(X|Z_y) - h(X|Y)
\end{equation}

In the case of lossless compression, $h(X|Z_y) = h(X|Y)$, so 

\begin{equation}
    I(X; Y) - I(X; Z_y) = h(X|Z_y) - h(X|Y) = 0
\end{equation}

However, if compression does not perfectly preserve information, it is possible that $h(X|Z_y) \gg h(X|Y)$. Since $h(X|Y)$ is intrinsic to the data and independent of learned representations, minimizing $h(X|Z_y)$ is equivalent to minimizing $I(X; Y) - I(X; Z_y)$. This points to an approach to learn representations suitable for approximating $I(X;Y)$: regularizing a pair of autoencoders to learn compressed representations $Z_x = f(X)$ and $Z_y = g(Y)$ while minimizing $h(X|Z_y)$ and $h(Y|Z_x)$. 

Because directly minimizing conditional entropies is intractable, we instead minimize a convenient proxy, which is the mean-squared error (MSE) loss of networks that predict one variable from another. The connection between conditional entropy and reconstruction loss has long been appreciated as a way to interpret autoencoders through an information-theoretic lens \cite{Vincent2008-ig, Devon_Hjelm2018-xq}. Here, we observe that this connection can be applied to learn representations which lend themselves to MI estimation. We explicitly show that minimizing cross-prediction loss from $Z_x$ to $Y$ is equivalent to minimizing an upper bound on the conditional entropy $h(Y|Z_x)$ in \textcolor{black}{Appendix A.1.1, Theorem 1}. 

Applying cross-predictive regularization to a pair of autoencoders results in a network architecture (Fig. 1) with one encoder for each variable and four decoders which reconstruct each variable from each  latent code. We train the networks by minimizing the sum of the MSE reconstruction loss for each decoder. More precisely, for variables $X, Y$ with dimensionality $d_X, d_Y$, we optimize encoders $E_X, E_Y$, and decoders $D_{XX}, D_{XY}, D_{YY}, D_{YX}$ to minimize $\mathcal{L}_{AEC} = \mathcal{L}_{AE} + \mathcal{L}_{C}$, where

\begin{equation}
    \mathcal{L}_{AE} = \frac{1}{d_X}\mathbb{E}[||X - D_{XX} (E_X(X))||^2_2] + \frac{1}{d_Y}\mathbb{E}[||Y - D_{YY} (E_Y(Y))||^2_2]
\end{equation}

\begin{equation}
    \mathcal{L}_{C} = \frac{1}{d_X}\mathbb{E}[||X - D_{YX} (E_Y(Y))||^2_2] + \frac{1}{d_Y}\mathbb{E}[||Y - D_{XY} (E_X(X))||^2_2]
\end{equation}

This is not the only way one could regularize autoencoders to preserve mutually informative structure. We design and empirically characterize some alternatives in \textcolor{black}{Appendix A.2}. We find that multiple regularization approaches can be effective, but cross-prediction comes with unique benefits. For example cross-predictive networks can be dissected to attribute high-dimensional MI estimates to specific dimensions, as demonstrated in \textcolor{black}{Appendix A.2.3.}

While the specific architecture of encoders and decoders could be carefully chosen for each estimation problem (e.g. convolutional layers for image data), here we use multilayer perceptrons with \textit{a priori} determined hidden layer sizes for all problems. This is intentional: a useful MI estimator should not need extensive parameter selection. Every LMI estimate shown in this paper (excluding Appendix) uses the default parameters of our library, equivalent to running \verb|lmi(X_samples,Y_samples)|.

To ensure that optimizing based on cross-reconstruction does not introduce spurious dependence due to overfitting, we learn representations and estimate MI on different subsets of the data. That is, for $N$ joint samples, we train the network using a subset of $N/2$ samples, then estimate MI by applying the estimator of \cite{Kraskov2004-sh} to latent representations of the remaining $N/2$ samples. A high-level overview of an MI estimate using LMI approximation is given in Algorithm 1.

We also state and prove some basic properties of LMI approximation, namely that $I(Z_x; Z_y) \leq I(X; Y)$, and that $I(Z_x; Z_y) = 0$ if $I(X;Y) = 0$ in \textcolor{black}{Appendix 1.3, Theorems 2 and 3}. 

\begin{algorithm}
    \caption{Estimating MI using LMI Approximation}
    \begin{algorithmic}
    \Require{$N$ joint samples $\{ (x_i, y_i) \}^N_{i=1}$ of random variables $X, Y$}
    \Require{Encoders $E_X, E_Y,$ decoders $D_{XX}, \ldots, D_{YY}$ parameterized by $\theta_1, \ldots, \theta_6$}

    \State randomly \textbf{split} into two subsets of $N/2$ samples, $\mathcal{D}_{\text{train}}, \mathcal{D}_{\text{est}}$

    \State \textbf{optimize} $\theta_1, \ldots, \theta_6$ to minimize $\mathcal{L}_{AEC}$ on $\mathcal{D}_{\text{train}}$

    \State \textbf{encode} $\mathcal{D}_{\text{est}}$ using $E_1, E_2, \theta_1, \theta_2$, yielding $\{ (Z^x_i, Z^y_i) \}_{i=1}^{N/2}$

    \State \textbf{return} $\hat{I}_{KSG} (\{ (Z^x_i, Z^y_i) \}_{i=1}^{N/2})$

    \end{algorithmic}
\end{algorithm}

\section{Empirical evaluation}

Next, we empirically study the effectiveness of LMI approximation with 16 dimensional latent space (8 dimensions per variable), in comparison with three popular estimators: the nonparametric estimator from \cite{Kraskov2004-sh}, referred to as KSG, and the variational bound estimators from \cite{Belghazi2018-ik, Van_den_Oord2018-lp} referred to as MINE and InfoNCE, respectively (implementation details in \textcolor{black}{Appendix A4}).

\subsection{Evaluating mutual information estimators on synthetic data}

We first consider the problem of MI estimation between multivariate Gaussian distributions, because ground truth MI can be analytically computed, and dimensionality can be easily tuned. We consider the scalability of MI estimators with increasing dimensionality of two kinds: the ambient dimensionality of the data, denoted $d$, and the intrinsic dimensionality of the dependence structure, denoted $k$. We benchmark the performance of estimators in the regime of high ambient dimensionality and low intrinsic dimensionality. Specifically, we consider variables with $d=10$ to $d=5 \cdot 10^3$ ambient dimensions and $k=1$ to $k=9$ dimensional dependence structure.

To generate samples from two $d$-dimensional random variables $X, Y$ with $k$-dimensional dependence structure, we sample $d$ bivariate Gaussians and concatenate the first components to construct samples of $X$, and concatenate the second components to construct $Y$. By choosing the covariance of each of the bivariate Gaussians, $I(X, Y)$ and $k$ can be tuned. To enforce a $k$-dimensional dependence structure, we can choose covariance matrices such that $\text{Cor}(X_i, Y_i) = \rho \textbf{Heav}(i-k)$ where $\textbf{Heav}$ is the Heaviside step function, and $\text{Cor}(X_i, Y_j)=0 \ \forall i \neq j$. The exact sampling procedure we use for these experiments is given in \textcolor{black}{Appendix A.3.1, Algorithm 4.}

Results of benchmarking MI estimators using these synthetic datasets are given in Fig. 2. For estimates from $N=2\cdot10^3$ samples, we find that, as expected, the performance of existing estimators degrades with $d$, with near complete failure for $d>100$ (Fig. 2a-c, 2f-h.). In contrast, applying LMI approximation results in stable estimates up to $d=5 \cdot 10^3$ ambient dimensions (Fig. 2d, 2i). The faithfulness of LMI approximation instead degrades with increasing $k$. Nonetheless, LMI approximation gives more absolutely and relatively accurate MI estimates than alternatives for 83\% and 87\% of tested settings respectively (Fig. 2e, 2j).

\begin{figure}
    \centering
    \includegraphics[width=\textwidth]{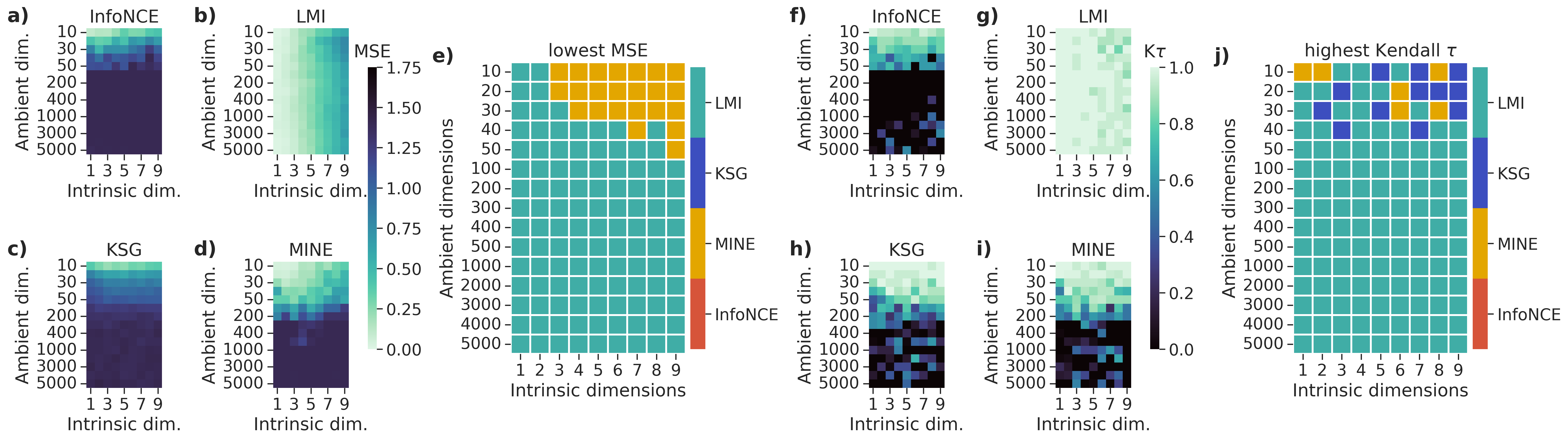}
    \caption{\textbf{MI estimator performance scaling with increasing dimensionality.} \textbf{a) - d)} Absolute accuracy measured by mean-squared error over 10 estimates per setting, with ground truth MI between 0 and 2 bits, and $2\cdot10^3$ samples per estimate. \textbf{e)} Estimator with highest absolute accuracy in each setting. Ties broken randomly. \textbf{f) - i)} Relative accuracy measured by Kendall $\tau$ rank correlation of estimates with ground truth. \textbf{j)} Estimator with highest relative accuracy in each setting. Ties broken randomly. }
    \label{fig:enter-label}
\end{figure}

\subsubsection{Empirically quantifying convergence rates of MI estimators on synthetic data}

The principle enabling the scalability of LMI approximation is that the number of samples it requires to converge is limited by $k$ rather than $d$ when $k \ll d$. We empirically demonstrate this by quantifying the convergence rates of MI estimators on the synthetic Gaussian datasets described above. We generate datasets with sample numbers in $N \in [10^2, 10^4]$, and ambient dimensionalities in $d \in [1, 50]$, each with a single correlated dimension between variables ($k=1$), and 1 bit MI. For each estimator and each ambient dimensionality $d$, we empirically determine the number of samples required to achieve $|I(X, Y) - \hat{I} (X, Y)| < \epsilon \text{ bits}$, with linear interpolation between tested sample numbers. 

As expected, methods that do not explicitly learn low-dimensional representations (InfoNCE, MINE, KSG) require increasing numbers of samples to estimate MI with error below $\epsilon = 0.1$ (Fig. 3a). KSG fails to estimate MI for $d \geq 13 $ for any $N$, while MINE and InfoNCE scale slightly better, failing for $d \geq 25$ and $d \geq 37$ respectively. In contrast, the sample requirements of LMI remain qualitatively stable -- no more than $4 \cdot 10^3$ samples are necessary for an accurate estimate.

While the convergence behavior of LMI is mostly unaffected by varying $d$, it is sensitive to varying $k$. When the same experiment is performed with increasing numbers of correlated dimensions at the limit where $k=d$, the convergence behavior of LMI is no longer favored over other estimators (Fig. 3c). The performance of all estimators dramatically decreases with $k$, such that a larger error tolerance must be chosen for informative convergence estimates. The dependence of variational bound estimator convergence on $k$ is, to our knowledge, not explained by existing theory \cite{Poole2019-iu, McAllester2020-vy}. In the intermediate case of $k = \lfloor 0.1 \cdot d \rfloor$ (Fig. 3b), we find that LMI convergence is fast with low $k$, but becomes slow as $k$ grows, nonetheless remaining favorable compared to other estimators.

\begin{figure}
    \centering
    \includegraphics[width=0.9\textwidth]{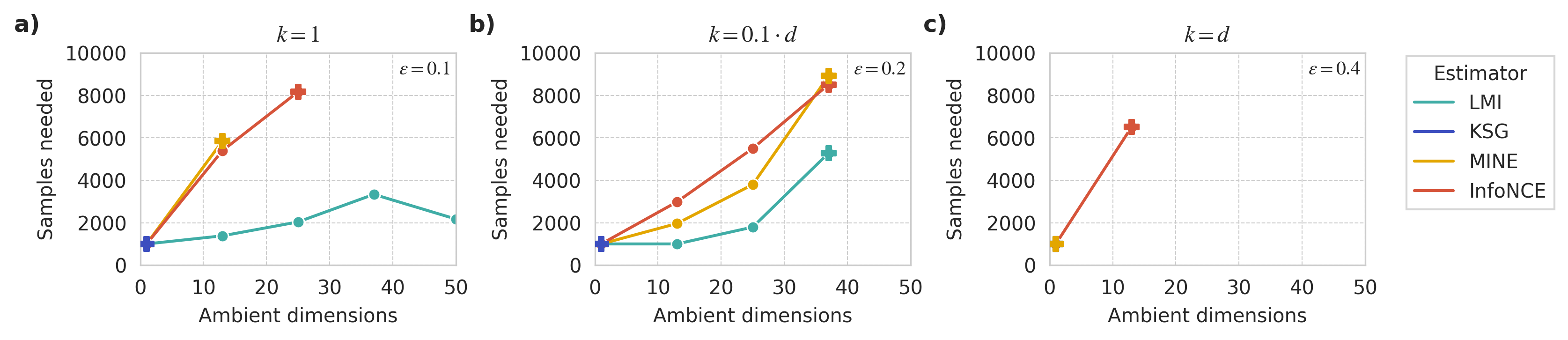}
    \caption{\textbf{Number of samples required to achieve $|I(X, Y) - \hat{I} (X, Y)| < \epsilon$.} \textbf{a) } Data with low-rank dependence structure, with $\epsilon = 0.1$. \textbf{b) } Moderate-rank dependence structure, with $\epsilon=0.2$. \textbf{c) } Full-rank dependence structure, with $\epsilon=0.4$  ``+'' marker indicates that $N>10^4$ samples are required for accurate estimates for all larger $d$.}
    \label{fig:enter-label}
\end{figure}

\subsection{Evaluating mutual information estimators on resampled real-world data}

While the empirical results on multivariate Gaussians are reassuring, they are not representative of performance on real data, where low intrinsic dimensionality is not known \textit{a priori}, and distributions can be non-Gaussian. To better understand the behavior of LMI in more realistic settings, we introduce a technique for creating benchmark datasets by resampling real-world data. Briefly, we use correspondences between discrete labels and complex data (i.e. digit labels and digit images in MNIST) to transform simple discrete distributions into realistic high-dimensional distributions.

Specifically, we draw samples from a bivariate Bernoulli vector, $\mathbf{L} = [L_x, L_y] \in \{0, 1\}^2$ with prescribed pairwise correlation $\text{Cor}(L_x, L_y) = \rho$, where each value corresponds to a discrete label of a set of samples in a high-dimensional dataset (e.g. $0$ and $1$ correspond to images of 0s and 1s in MNIST). For each sample of $\mathbf{L}$, we replace each component with a random (without replacement) high-dimensional sample matching the label. For the example of MNIST, this transforms samples from $\mathbf{L}$ into pairs of images of 0s and 1s, represented as samples of random vectors $X, Y \in \mathbb{R}^{784}$.

Under the assumption that discrete labels can be uniquely identified by high-dimensional vectors, high-dimensional MI is identical to the discrete label MI. That is, assuming $H(L_x | X) = H(L_y | Y) = 0$ then $I(X;Y) = I(L_x;L_y)$ (shown in \textcolor{black}{Appendix A.3.2, Theorem 4}). And using our knowledge of $\rho$, $I(L_x; L_y)$ can be analytically computed. 

We resample two different source datasets: (1) ``binary'' subset of MNIST, containing only images of 0s and 1s, with $5000$ samples and $784$ dimensions and (2) embeddings of a subset of protein sequences from \textit{E. coli} and \textit{A. thaliana} proteins, with $4402$ samples and $1024$ dimensions. For both source datasets, we validate the $I(L_x; L_y) \approx I(X;Y)$ approximation in \textcolor{black}{Appendix A.3.3}. 

For each source dataset, we generate 200 benchmark datasets with true MI ranging from 0 to 1 bits. We estimate MI on each dataset with each estimator (Fig. 4a, 4b), and quantify absolute accuracy (MSE), relative accuracy (rank correlation with ground truth), and runtime for each estimator (Fig. 4c). For both types of source data, we find that variational bound estimators have high variance, in line with previous observations \cite{Poole2019-iu}. On protein embedding datasets, variational estimators nearly always fail to estimate nonzero values -- resulting in a rank correlation below 0 for InfoNCE. The KSG estimator, while achieving high relative accuracy, systematically underestimates MI, resulting in low absolute accuracy. Furthermore, the amount by which it underestimates true MI is different between the two datasets -- indicating inequitability. In contrast, LMI approximation yields estimates consistently close to the ground truth, with high relative and absolute accuracy.

\begin{figure}
    \centering
    \includegraphics[width=\textwidth]{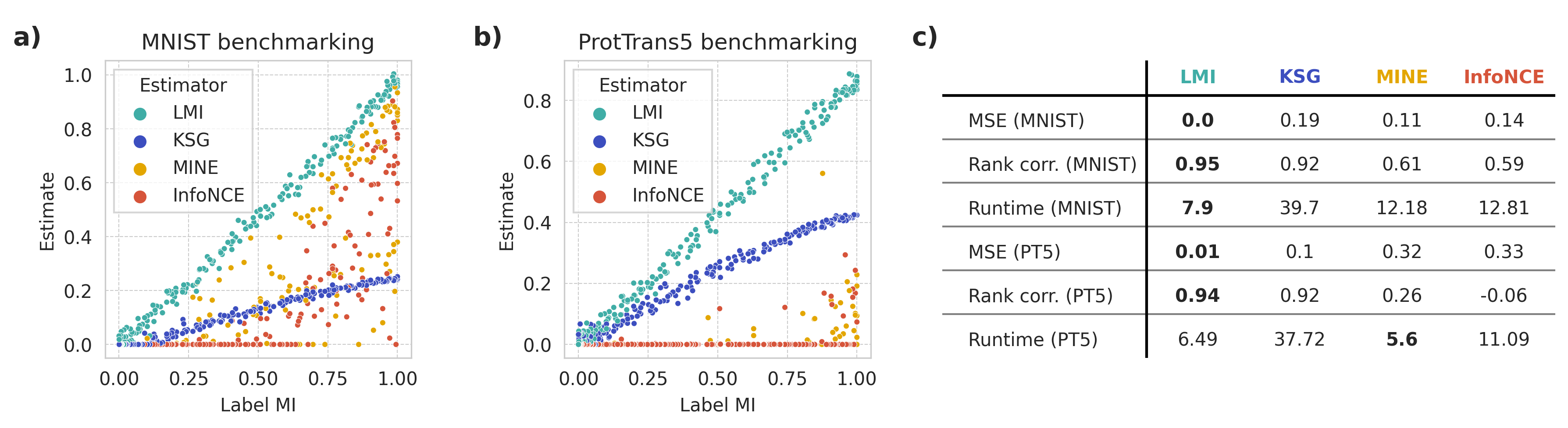}
    \caption{\textbf{Performance of MI estimators on resampled real datasets.} \textbf{a)} Estimates on resampled pairs of MNIST digits, with $5\cdot10^3$ samples and $784$ dimensions. \textbf{b)} Estimates on resampled pairs of ProtTrans5 sequence embeddings, with $4.4 \cdot 10^3$ samples and $1024$ dimensions. \textbf{c)} Statistics of estimator accuracy and runtime (in seconds), for each dataset type.}
    \label{fig:clusters}
\end{figure}

\section{Applications}

\subsection{Quantifying interaction information in protein language model embeddings}

\begin{figure}
    \centering
    \includegraphics[width=\textwidth]{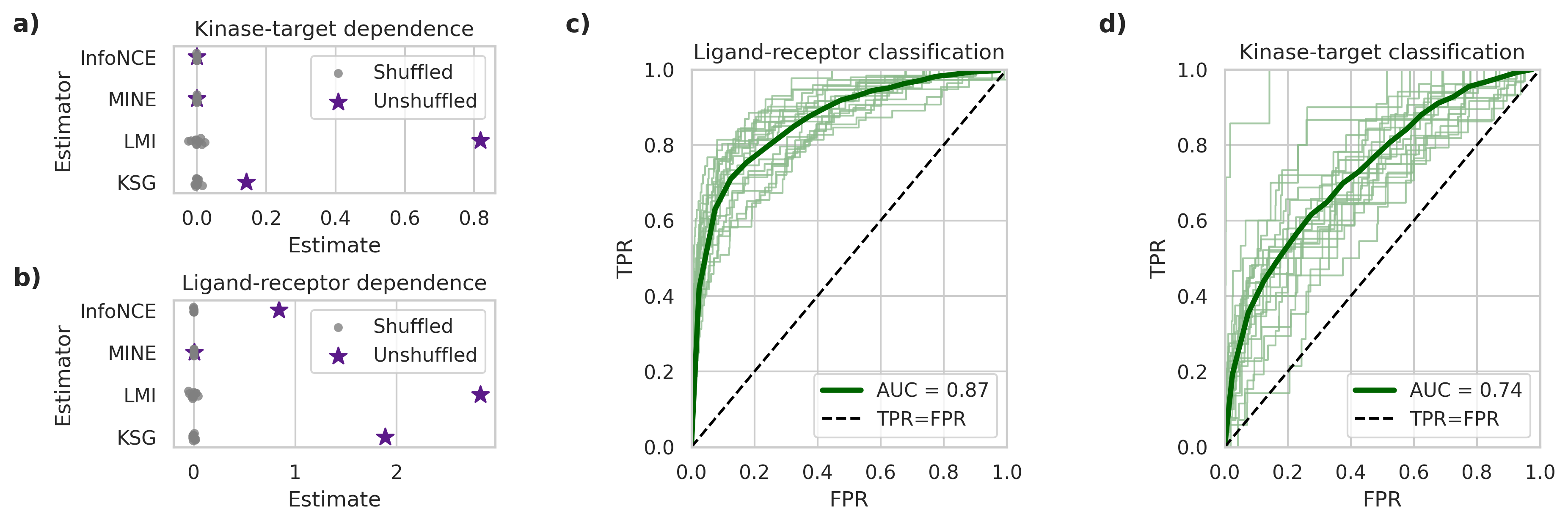}
    \caption{\textbf{Quantifying dependence between participants of protein interactions.} \textbf{a) - b)} MI estimates between interaction partners, compared to randomly permuted data. \textbf{c) - d)} ROC curves of density ratio classifier distinguishing annotated interacting pairs from unannotated ``negative'' samples, for all pairs of $170$ held-out proteins. Averages over 20 random hold-out splits.}
    \label{fig:enter-label}
\end{figure}

Pretrained protein language models (pLMs) have recently seen widespread use, largely due to their convenient representations of protein sequences (vectors in $\mathbb{R}^N$, typically with $N \approx 10^3$) which can be used for transfer learning on downstream tasks \cite{Rives2021-ws, Elnaggar2020-eq, Alley2019-bx}. While it is known that pLM sequence embeddings contain significant information about protein structure \cite{Zhang2024-li}, it is not clear how well existing pLMs encode functional information. Recent work has shown that pLMs fail to capture some important functional properties, such as thermostability \cite{Li2024-vd}. It remains unclear the extent to which pLM embeddings contain information about protein-protein interactions, which are essential to protein function. Here, we use an information-theoretic approach to quantify interaction information contained in 1024-dimensional sequence embeddings from the ProtTrans5 model \cite{Elnaggar2020-eq}. 

We study two types of protein-protein interactions: kinase-target and ligand-receptor interactions. For both, the OmniPath database \cite{Turei2021-ur} contains lists of thousands of annotated pairs of interacting proteins. We consider each annotated pair to be a sample from a joint ``interaction'' distribution over pairs of sequence embeddings. For example, for kinase-target interactions, we consider kinase and target sequence embeddings as random variables $K \in \mathbb{R}^{1024}, T \in \mathbb{R}^{1024}$, with joint distribution $P_{KT}(k, t)$. Then, using joint samples $\{(k_1, t_1), \ldots, (k_N, t_N)\}$ we estimate mutual information between interaction partners, $I(K;T)$. We analogously estimate $I(L;R)$ for ligand-receptor interactions.

If pLM embeddings capture interaction information, MI between interaction partners should be significantly above 0 bits. Applying LMI approximation, we estimate $ I(L;R) \approx 2.8 \text{ bits}$ and $I(K;T) \approx 0.8 \text{ bits}$. To test the significance of these values, we estimated MI from shuffled data, and found $I_{\text{shuff}}(K; T) \approx I_{\text{shuff}}(L; R) \approx 0$ (both mean and standard deviation $<0.05$), across 20 random shuffles. These results indicate that pLM embeddings contain information about both types of interactions, and contain more information about ligand-receptor interactions than kinase-target interactions. In contrast, existing estimators yield far lower estimates, with MINE estimates indicating independence for both types of interactions (Fig. 5a, 5b). To validate the LMI estimates of dependence, we next operationally verify the presence of interaction information.

If protein-protein interactions can be predicted for a set of held-out proteins based on sequence embeddings, then sequence embeddings must contain interaction information. To see if this is the case, we extend LMI to predict protein interactions from sequence embeddings. For ligand-receptor prediction, our goal is to predict whether a held-out pair of sequence embeddings $(l, r)$ is an annotated ligand-receptor pair. One way to do this is estimating the log density ratio $\log \frac{P_{LR}(l, r)}{P_L(l)\cdot P_R(r)}$, and setting a threshold above which sequence pairs are predicted to be annotated interactions. 

We make a simple modification to the KSG estimator to yield the desired density ratio estimates (given in \textcolor{black}{Algorithm 2}), and use these estimates (with latent approximation) to predict interaction annotations. For 20 different random splits of 170 held out proteins, we use density ratio estimates to classify all $2.89\cdot10^4$ pairs of held out proteins as interacting or non-interacting. The receiver operating characteristic (ROC) curves for predictions of both interaction types are shown in Fig. 5c, 5d, with mean AUC-ROC scores of 0.87 and 0.74 for kinase-target and ligand-receptor interactions respectively. These results demonstrate that protein interactions can be predicted better than random chance using ProtTrans5 embeddings, suggesting that pLM embeddings capture information about kinase-target and ligand-receptor interactions. And in line with the LMI estimates, ligand-receptor interactions are better predicted than kinase-target interactions.

\begin{algorithm}
    \caption{k-nearest neighbor log density ratio estimator}
    \begin{algorithmic}
        \Require joint samples $\{(x_i, y_i)\}_{i=1}^N$
        \Require query point $(q_x, q_y)$
        \State let  $(r_x, r_y) \gets \ $ $k$-th nearest neighbor sample of $(q_x, q_y)$ in joint space (default $k=3$)
        \State \textbackslash \textbackslash  \ compute Chebyshev distance
        \State let $d \gets ||(q_x, q_y) - (r_x, r_y)||_{\infty}$
        \State let $n_x \gets 0, n_y \gets 0$
        \State \textbackslash \textbackslash  \ count neighbors within $d$ in marginal spaces
        \For{each $(x_i, y_i)$}
        \If{$||q_x - x_i||_{\infty} < d$}
        \State $n_x$ += 1
        \EndIf
        \If{$||q_y - y_i||_{\infty} < d$}
        \State $n_y$ += 1
        \EndIf
        \EndFor
        \State \textbackslash \textbackslash  \ return estimate of $\log \frac{p(q_x, q_y)}{p(q_x) p(q_y)}$
        \State \textbf{return} $\psi(k) + \psi(N) - \psi(n_x) - \psi(n_y)$, where $\psi$ is Digamma function
    \end{algorithmic}
\end{algorithm}

\subsection{Identifying cell fate information in hematopoietic stem cells}

Single cell RNA sequencing (scRNA-seq) measures the expression of $g \approx 10^4$ genes in single cells, which can be thought of samples of a gene expression state variable $X \in \mathbb{R}^g$. These samples can be used to infer a probability distribution over gene expression states, $P_{X}$. To study the dynamics of gene expression, one approach is to make measurements at multiple timepoints $t_1, ... t_N$, which can be thought of as samples of random variables $X_{i} \in \mathbb{R}^g$. Lineage tracing is a technique where clonally related cells can be labelled with barcodes, allowing sampled cells from different timepoints to be matched with their ``twins'' from another. When combined with scRNA-seq, lineage tracing can be thought to provide samples from the joint distribution $P_{X_{1},\ldots,X_{N}}$ \cite{Wang2022-bg}.

One fundamental question about cellular dynamics is whether the time evolution of gene expression state is dependent entirely on the current gene expression state. That is, if the ``fate'' of a cell can be formally modeled as a Markov chain $X_{i} \to X_{i+1}$. In some cases, cell behavior may be a function of hidden variables resulting in non-Markovian dynamics. Using the data processing inequality (DPI) \cite{Cover2006-hn}, we know that if gene expression dynamics are Markovian, $I(X_{i}; X_{i+1}) \geq I(X_{i}; X_{i+2})$, and if the DPI is does not hold, then gene expression states are non-Markovian. This can, in principle, be tested using samples from the joint distribution $P_{X_{1},..,X_{N}}$. Due to the difficulty of high-dimensional MI estimation, previous work has used heuristic alternatives to indirectly test the Markovian assumption for lineage-traced scRNA-seq (LT-seq) data \cite{Weinreb2020-uk}. Here, we will use LMI to explicitly estimate high-dimensional MI from LT-seq data, and test the Markovian assumption for gene expression states.

We study a previously published LT-seq data set of \textit{in vitro} differentiating mouse hematopoietic stem cells \cite{Weinreb2020-uk}. The dataset includes sister cells which are separated at day 2 of the experiment and allowed to differentiate in separate wells until day 6, with cell states sampled on both days 2 and 6. Under the Markovian assumption, this can be modeled as $X_2 \to X_6$, $X_2 \to X_{6'}$ and by the DPI we should have $I(X_2; X_6) \geq I(X_6; X_{6'})$. Using LMI, we estimate that $I(X_2; X_6) \approx 0.31 \pm 0.02 \text{ bits}$ and $I(X_6; X_{6'})\approx 0.98 \pm 0.01 \text{ bits}$ (mean $\pm$ SEM), over 20 random pairings of clonally related cells. This indicates that gene expression states are non-Markovian, in line with prior findings \cite{Weinreb2020-uk, Jindal2023-dl}.

Cell fate information manifests in transcriptomes sometime between days 2 and 6. By decomposing LMI estimates into pointwise contributions \textcolor{black}{(Appendix A.4.3, Algorithm 5)}, we can determine precisely when this hidden information emerges. Generally, we see that pMI increases along differentiation trajectories (Fig. 6b). Along the neutrophil trajectory, we quantatively compare cell fate information with neutrophil pseudotime computed by graph smoothing \cite{Weinreb2020-uk}, and find that pMI begins to rapidly increase around pseudotime value of $35\cdot10^3$, which is roughly aligned with the transition from granulocyte-myeloid progenitor to promyelocyte, as defined in \cite{Weinreb2020-uk}.

\begin{figure}
    \centering
    \includegraphics[width=\textwidth]{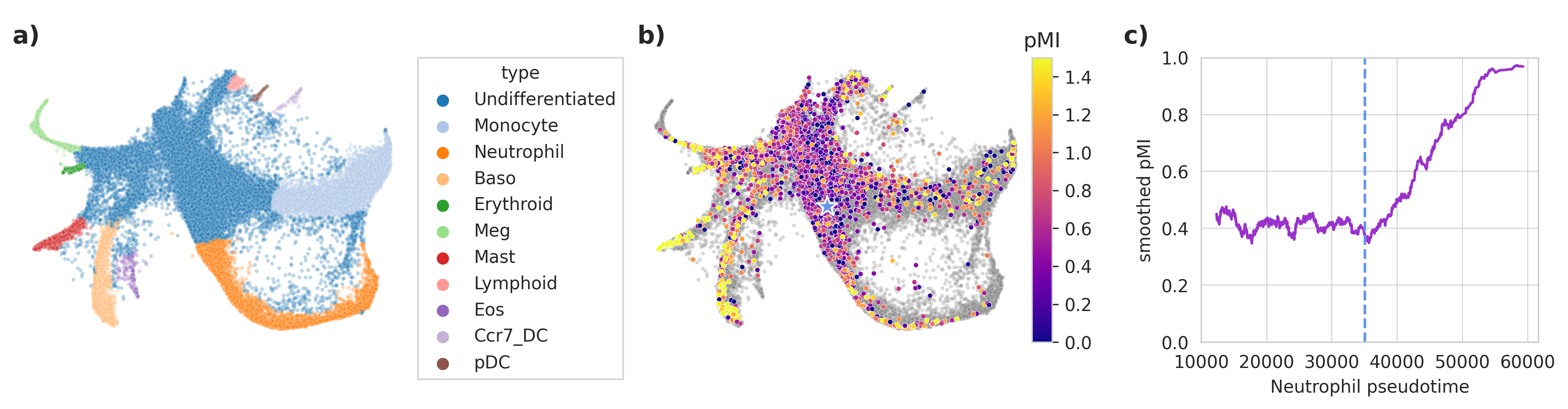}
    \caption{\textbf{Quantifying cell fate information in single cell transcriptomes.} \textbf{a)} 2D SPRING embedding \cite{Weinreb2018-sp} of lineage-traced single cell RNA-sequencing data from \cite{Weinreb2020-uk}. \textbf{b)} Pointwise decomposition of MI between sister cells across timepoints, as estimated using LMI. Hue applied to early timepoint cells, late timepoint and unbarcoded cells in grey. Star indicates neutrophil pseudotime value $35\cdot10^3$. \textbf{c)} Smoothed pMI (rolling average over 100 cells) across neutrophil differentiation trajectory. Vertical line denotes neutrophil pseudotime value $35\cdot10^3$.}
    \label{fig:hematopoiesis}
\end{figure}
\section{Discussion}

In this paper, we tested the hypothesis that low-dimensional structure can enable scalable estimation of MI from high-dimensional data. We introduced LMI approximation, which applies a nonparametric MI estimator to low-dimensional representations learned by neural networks. We quantified the effectiveness of  LMI approximation, using multiple approaches spanning over 3000 benchmark datasets. Our results suggest that, unlike existing techniques, LMI approximation is generally effective for high-dimensional data with low-dimensional structure, even if the number of available samples remains relatively low -- a regime where many real datasets reside \cite{Thibeault2024-an}.

We used LMI to study two open problems in biology. We show one example where LMI enables the use of information-theoretic ideas to study the dynamics of gene expression. In the original study \cite{Weinreb2018-sp}, the authors had indeed wished to estimate MI but were unable to do so and resorted to heuristic approaches. LMI may similarly help identify dependence in cellular dynamics in other systems \cite{Jindal2023-dl, Chadly2024-ga, Wagner2020-xr}. In a different subfield of biology, we showed an example of using LMI to quantify functional information learned by pLMs. Our results suggest that nontrivial protein-protein interaction information is learned by ProtTrans5, motivating the development of interaction prediction tools based on pLMs. As the number of large pLMs grows \cite{Rives2021-ws, Alley2019-bx, Lin2023-nr, Elnaggar2020-eq}, information-theoretic approaches using LMI could help benchmark and compare models. 

More broadly, LMI may be able to help bridge the gap between theory and experimental measurements in the study of complex systems \cite{Bruckner2023-yh, Tkacik2016-uo, Bray1995-dp, Shwartz-Ziv2017-us}. With the goal of making high-dimensional MI estimation accessible to practitioners, we provide an open-source implementation of LMI approximation along with the code necessary to reproduce the results of this paper at \href{https://anonymous.4open.science/r/latent-mutual-information-BCBF/README.md}{this link}. 

\paragraph{Limitations} The most prominent limitation of LMI follows directly from its motivating assumption: it will fail to capture dependence structure with dimensionality larger than its learned representation. As a result, it is very easy to design synthetic datasets on which LMI will fail entirely (see Fig. 3c). However, it is likely that many real datasets (beyond those explored in this paper) will be amenable to LMI approximation, as there is strong evidence that complex systems generally have low intrinsic dimensionality \cite{Thibeault2024-an}. Our implementation of LMI approximation also inherits some limitations of the KSG estimator, notably that it fails for strongly dependent (near deterministically related) variables \cite{McAllester2020-vy, Gao2014-tb}. To overcome this, it may be possible to apply previously developed corrections \cite{Gao2014-tb}. Finally, despite reassuring empirical results, few theoretical properties of LMI have been derived. This is an important line of future work, which could help precisely identify settings where  LMI approximation is effective. 

\paragraph{Broader impacts} MI estimators have been used to quantify moral and legal fairness \cite{Chen2023-nj, Cho2020-cx, Kang2022-ri}. LMI approximation is not universally faithful, and more generally no MI estimator can be universally accurate \cite{McAllester2020-vy}. MI estimates must be interpreted with great care when applied to human lives. The experiments (and pilot iterations) in this paper were performed on a single NVIDIA RTX 3090, and resulted in estimated 45.36 kg CO$_2$eq. Estimates made using \cite{Lacoste2019-cz}. 

\paragraph{Code reproducibility} The \verb|lmi| Python library and the code necessary to reproduce the results from this paper are available at \url{https://anonymous.4open.science/r/latent-mutual-information-BCBF}. Documentation for the \verb|lmi| library can be found at \url{https://latentmi.readthedocs.io/en/latest/}.

\paragraph{Author contributions} G.G. conceived of, designed, and performed the study, analyzed the data, and wrote the paper. X.L. contributed to study design. A.M.K. contributed to the design of the study, analysis of the data, and wrote the paper, and supervised the study. P.Y. supervised the study. All authors reviewed, edited, and approve the paper.

\begin{ack}
We thank Caroline Holmes, Sean McGeary, and Pippa Richter for thoughtful discussions. This work is supported by funding from NIH Pioneer Award DP1GM133052, R01HG012926 to P.Y., and Molecular Robotics Initiative at the Wyss Institute.
\end{ack}

\bibliographystyle{unsrt}
\bibliography{c}

\clearpage
\appendix

\section{Appendix / supplemental material}

In this appendix, we first elaborate on the theory and implementation details of the LMI approximation. Then, we will discuss some alternative approaches one could use to implement the LMI approximation. Then, we describe the details, assumptions, and motivations of our empirical evaluation benchmarks. Finally, we provide details relevant to reproducibility -- specifically, our implementations of existing estimators and data preprocessing methods (all of which are also included in our code supplement).

\subsection{Theory and implementation of LMI}

\subsubsection{Theoretically motivating the cross-predictive representation learning architecture}

The core theoretical underpinning of the network architecture used in the LMI approximation is that cross-predictive mean-squared loss is a proxy of conditional entropy. Here, we explicitly show this.

\begin{theorem}
Let $X = [X_1,\ldots,X_d]$ and $Z = [Z_1,\ldots,Z_k]$ be absolutely continuous random vectors in $\mathbb{R}^d$ and $\mathbb{R}^k$ respectively with finite differential entropy. Let $f_\theta : \mathbb{R}^k \to \mathbb{R}^d$ be a function (a neural network parameterized by $\theta$) to estimate $\hat{X} = f_\theta(Z)$. For any $\theta$,

\begin{equation}
    h(X|Z) \leq \alpha + \frac{1}{2}\log \text{MSE}(\hat{X}, X) 
\end{equation}

where $\alpha$ is a positive constant and $\text{MSE}(\hat{X}, X) = \frac{1}{d} \sum_i \mathbb{E}[(X_i - \hat{X}_i)^2]$
\end{theorem}

\begin{proof}
From the chain rule for differential entropy, we can bound
 
\begin{equation}
    h(X|Z) \leq \sum_i h(X_i | Z)
\end{equation}

Because the maximum entropy distribution with fixed variance is Gaussian \cite{Cover2006-hn}, we can bound

\begin{equation}
    \sum_i h(X_i | Z) \leq \sum_i \frac{1}{2} \log (2 \pi e \ \text{Var}(X_i | Z))
\end{equation}

\begin{equation}
    = \sum_i \frac{1}{2} \log (2 \pi e \ \mathbb{E}\big[(X_i - \mathbb{E}[X_i|Z])^2\big])
\end{equation}

Because the expectation of a random variable is its best estimator, \cite{Cover2006-hn}

\begin{equation}
    \sum_i \frac{1}{2} \log (2 \pi e \ \mathbb{E}\big[(X_i - \mathbb{E}[X_i|Z])^2\big]) \leq \sum_i \frac{1}{2} \log (2 \pi e \ \mathbb{E}\big[(X_i - \hat{X}_i\big)^2])
\end{equation}

So with positive constant $\alpha$,

\begin{equation}
    h(X|Z) \leq \alpha + \frac{1}{2}\log \text{MSE}(\hat{X}, X) 
\end{equation}
    
\end{proof}

 In LMI, $Z$ corresponds to the latent representation of one input variable ($Y$), and $f$ corresponds to the decoder which aims to reconstruct the other variable $X$ from the latent code $Z$.This result is very similar to some used in information-theoretic interpretations of autoencoders \cite{Vincent2008-ig, Devon_Hjelm2018-xq}, and can be thought of as a continuous analog of Fano's inequality \cite{Cover2006-hn}. The $d=k=1$ case of this bound is given as a Corollary to Theorem 8.6.6 in \cite{Cover2006-hn}.

 \subsubsection{Implementation of the representation learning architecture}

 There are many ways one could implement the high-level network architecture suggested by Theorem 1, with different encoder and decoder architectures, and choices of hyperparameters. Here, we will describe our design choices. Our motivating philosophy was that the implementation details should not need to be tuned for specific estimation problems. 

 All LMI estimates presented in the main text use programatically determined parameters, requiring no user input beyond joint samples. By default, all latent representations have 16 dimensions (8 for each variable). Each encoder and decoder is a multilayer perceptron (MLP) with three hidden layers, whose sizes are determined as a function of the dimensionality of the input variable. For a variable with dimensionality $d$, the encoder has hidden layer sizes $L, L/2, L/4$ with $L = \max (2^{\lfloor log_2(d) \rfloor}, 1024)$. Decoders have the same structure inverted, with hidden layer sizes $L/4, L/2, L$. 

All MLP activations used are Leaky ReLUs, with negative slope 0.2, except the last layers of decoders, which have no activation. Cross-decoders are trained with 50\% dropout after each activation layer. All weights are initialized using Xavier uniform initialization \cite{Glorot2010-dg}, and optimized using Adam, with hyperparameters listed in Table 1. They are trained with batch size of 512, with $1:1$ train-validation splits, and a maximum of 300 epochs using early stopping procedure provided in \textcolor{black}{Algorithm 3}.

All models are implemented in Pytorch \cite{Paszke2019-bf}. All experiments in this paper were done using a single NVIDIA RTX 3090.

\begin{table}[H]
\centering
\begin{tabular}{|c|c|}
\hline
\textbf{Parameter} & \textbf{Value} \\
\hline
Learning rate ($\alpha$) & $10^{-4}$ \\
\hline
$\beta_1$ & $0.9$ \\
\hline
$\beta_2$ & $0.999$ \\
\hline
Epsilon ($\epsilon$) & $10^{-7}$ \\
\hline
\end{tabular}
\caption{Adam optimizer parameters used in LMI.}
\end{table}

\begin{algorithm}
\caption{Early Stopping}
\begin{algorithmic}
\Procedure{EarlyStopping}{$\text{model}, \text{validation\_losses}, \text{patience}=30$}
    \State $\text{best\_loss} \gets \infty$
    \State $\text{patience\_counter} \gets 0$
    \State $\text{best\_model} \gets \text{None}$
    \While{$\text{patience\_counter} < \text{patience}$}
        \State $\text{current\_loss} \gets \text{validation\_loss}(\text{model})$
        \If{$\text{current\_loss} < \text{best\_loss}$}
            \State $\text{best\_loss} \gets \text{current\_loss}$
            \State $\text{patience\_counter} \gets 0$
            \State $\text{best\_model}\gets \text{model}$
        \Else
            \State $\text{patience\_counter} \gets \text{patience\_counter} + 1$
        \EndIf
    \EndWhile
    \State \textbf{return} $\text{best\_loss}$,  $\text{best\_model}$
\EndProcedure
\end{algorithmic}
\end{algorithm}

\subsubsection{Theoretical properties of LMI approximation}

The error of MI estimates using LMI approximation can be broadly attributed to two sources: error due to the representation approximation (i.e. $|I(X; Y) - I(Z_x; Z_y)|$), and classical MI estimation error (i.e. $|\hat{I}_{KSG}(Z_x; Z_y) - I(Z_x; Z_y)|$). Here, we will explore the first source by deriving some basic properties of the representation approximation $I(Z_x; Z_y)$.

\begin{theorem}
    Let $X, Y$ be random vectors in $\mathbb{R}^d$ and $Z_x = f_\theta(X)$, $Z_y = g_\phi(Y)$ where $f,g$ are neural networks parameterized by $\theta, \phi$. For any $\theta, \phi$,

    \begin{equation}
        I(Z_x; Z_y) \leq I(X; Y)
    \end{equation}
\end{theorem}

\begin{proof}
    Because $Z_y = f_\theta (Y)$, we have that $X \to Y \to Z_y$ form a Markov chain. From the data processing inequality, \cite{Cover2006-hn}

    \begin{equation}
        I(X; Y) \geq I(X; Z_y)
    \end{equation}

    Similarly, we have $Z_y \to X \to Z_x$, and

    \begin{equation}
        I(X; Y) \geq I(X; Z_y) \geq I(Z_x; Z_y)
    \end{equation}
    
\end{proof}

\begin{theorem}
    Let $X, Y$ be independent random vectors in $\mathbb{R}^d$, such that $I(X; Y) = 0$. Let $Z_x = f_\theta(X)$, $Z_y = g_\phi(Y)$ where $f,g$ are neural networks parameterized by $\theta, \phi$. For any $\theta, \phi$,

    \begin{equation}
        I(Z_x; Z_y) = 0
    \end{equation}
\end{theorem}

\begin{proof}
    Due to the nonnegativity of mutual information, we know

    \begin{equation}
        I(Z_x; Z_y) \geq 0
    \end{equation}

    From \textcolor{black}{Theorem 2}, we have

    \begin{equation}
        I(Z_x; Z_y) \leq I(X; Y) = 0 
    \end{equation}

    So,

    \begin{equation}
        0 \leq I(Z_x; Z_y) \leq 0
    \end{equation}

    \begin{equation}
        I(Z_x; Z_y) = 0
    \end{equation}
\end{proof}

\clearpage
\subsection{Alternate approaches to latent MI approximation}

The broad goal of LMI, to estimate $I(X;Y)$ using $I(Z_y; Z_x)$ where $Z_x, Z_y$ are low-dimensional representations, is quite general and could be approached in many ways beyond cross-predictive regularization. Next, we will discuss some alternative approaches, empirically explore their performance, and finally, show one unique advantage of the cross-predictive representation learning architectures.

\subsubsection{Alternate methods of regularizing autoencoders for MI estimation}

Another approach to learn $Z_x, Z_y$ such that $\hat{I}(Z_x; Z_y) \approx I(X; Y)$ is to regularize autoencoders to maximize $I(Z_x; Z_y)$. This approach is sensible because the data processing inequality ensures that $I(Z_x;Z_y) \leq I(X;Y)$. So maximizing $I(Z_x;Z_y)$ is equivalent to minimizing the approximation error $I(X;Y) - I(Z_x; Z_y)$.

While maximizing directly $I(Z_x; Z_y)$ is intractable, we can build on the variational bounds explored in \cite{Belghazi2018-ik, Van_den_Oord2018-lp}. We can add loss term $\mathcal{L}_{\text{MINE}} (Z_x; Z_y)$ or $\mathcal{L}_{\text{InfoNCE}} (Z_x; Z_y)$ to our autoencoder loss functions to regularize latent codes to preserve mutually informative structure.

We implement both of these approaches, and benchmark them using the approach described in \textcolor{black}{Figure 2 and Section 3.1}. We find that the estimates from these regularization approaches perform similarly, but slightly more poorly, than the cross-predictive regularization. 

\begin{figure}
     \centering
     \begin{subfigure}[b]{0.45\textwidth}
         \centering
         \includegraphics[width=\textwidth]{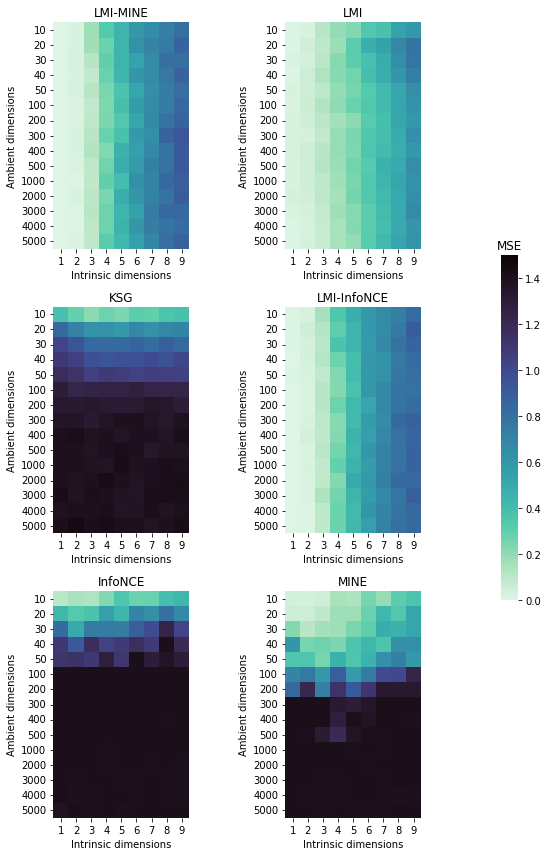}
         \caption{Absolute error, measured by MSE.}
         \label{fig:mses}
     \end{subfigure}
     \hfill
     \begin{subfigure}[b]{0.45\textwidth}
         \centering
         \includegraphics[width=\textwidth]{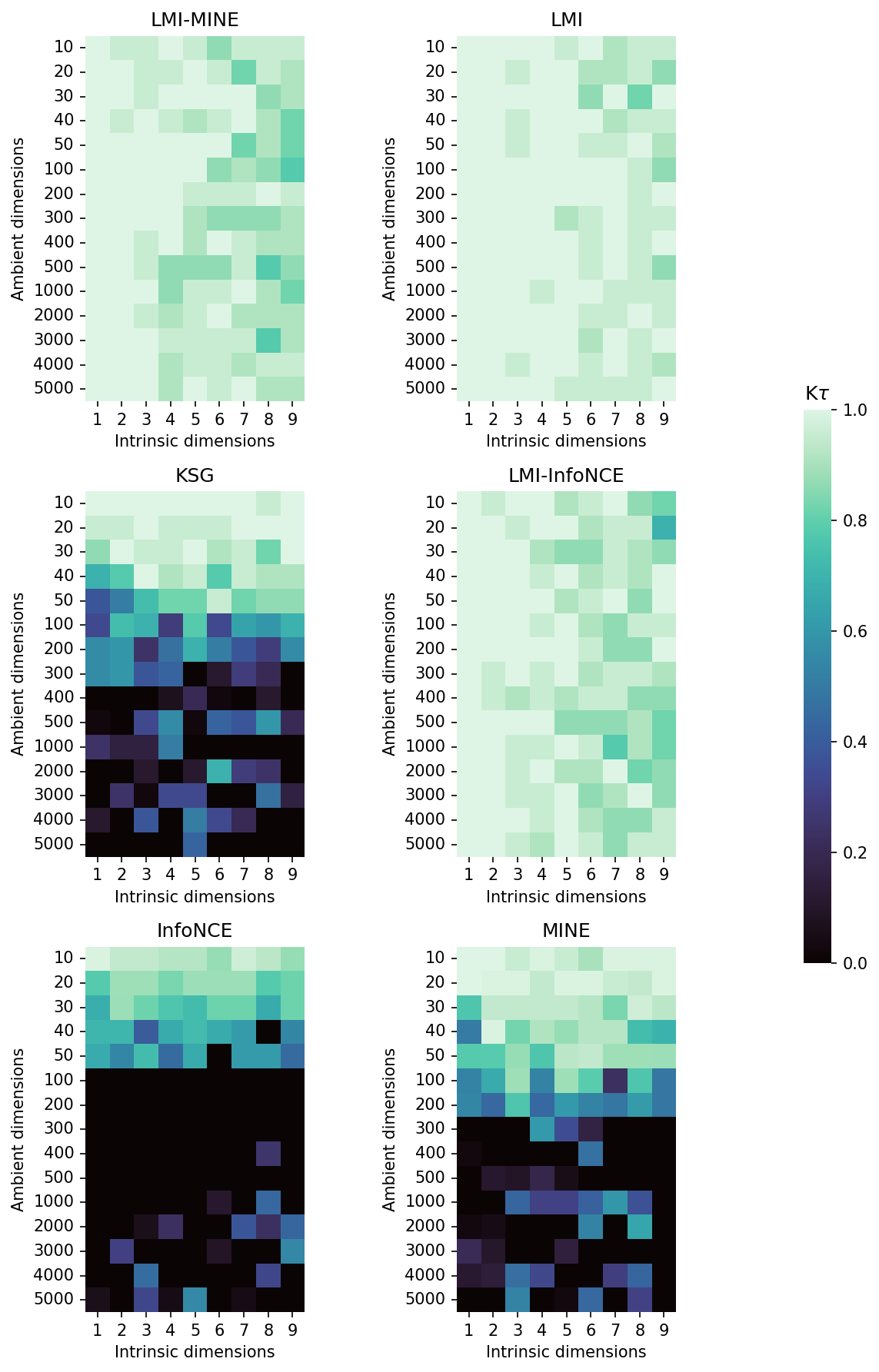}
         \caption{Relative error, measured by Kendall $\tau$.}
         \label{fig:kendalls}
     \end{subfigure}
     \caption{Experiment from Figure 2, with alternate regularization approaches.}
\end{figure}

\subsubsection{Comparing latent nonparametric and latent variational MI estimation}

After learning low-dimensional representations, there are multiple estimators that could be used for latent MI approximation. This appendix evaluates two methods: the KSG nonparametric estimator or the InfoNCE variational estimator. Nonparametric nearest-neighbor estimators have several advantages in low-dimensional settings: they generally require far fewer samples for accurate estimation \cite{Czyz2023-qs}, and yield pointwise mutual information decompositions (see \textcolor{black}{Algorithm 5}).

For completeness, we empirically compare both options here on one particular estimate. For a Gaussian dataset generated by \textcolor{black}{Algorithm 4} with $d=100$, $k=1$, $N=5\cdot10^3$ and 1 bit MI, we train autoencoders regularized by InfoNCE loss to maximize $I(Z_x; Z_y)$. After each training epoch, we measure LMI as estimated using latent KSG, and using latent InfoNCE. Both are plotted for 100 trials in \textcolor{black}{Appendix Figure 9}. The latent KSG estimation converges quickly to the true value, while the latent InfoNCE estimate converges somewhat slowly to a value below the ground truth.

\begin{figure}
    \centering
    \includegraphics[width=0.4\textwidth]{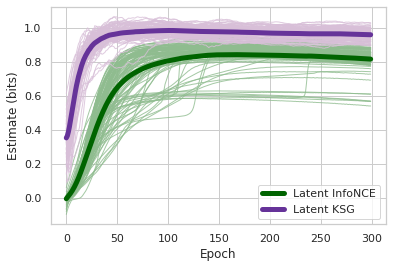}
    \caption{Convergence of multiple latent estimation approaches during training. Bold lines indicate averages over 100 trials. Ground truth is 1 bit MI.}
    \label{fig:enter-label}
\end{figure}

\subsubsection{Interpreting decoders with element-wise reconstruction error}

Beyond performance differences, one benefit of using cross-predictive networks to regularize latent representations is that the cross-decoders themselves are useful. For example, by inspecting the dimension-wise reconstruction error of decoders, we can attribute an MI estimate to the predictability of certain dimensions. For a network trained on the ``binary'' MNIST dataset with $L_x = L_y$, visualizing the dimension-wise reconstruction error of a cross-predictive decoder reveals the pixels that contain information about digit identity \textcolor{black}{(Fig. 10)}.

Pixels with low reconstruction error are likely to be ``well-explained'' by the other high-dimensional variable, while pixels with high reconstruction error are poorly explained. However, this reasoning is not universally applicable. Dimensions with no variation do not contribute to MI, but have very low reconstruction error. In \textcolor{black}{Fig. 10}, the outermost pixels are examples of this.

\begin{figure}
    \centering
    \includegraphics[width=0.4\textwidth]{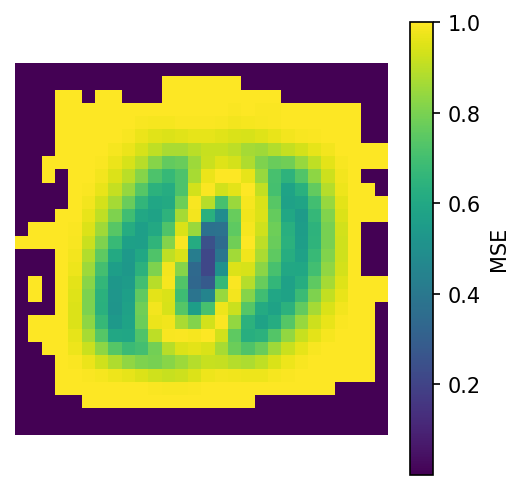}
    \caption{Pixel-wise reconstruction error of cross-decoders in paired binary MNIST dataset where $L_x = L_y$.}
    \label{fig:enter-label}
\end{figure}

\clearpage
\subsection{Details of experimental evaluation benchmarks}

In this section, we will describe the details of our experimental evaluation benchmarks from \textcolor{black}{Section 3}. First, we will describe how we generate multivariate Gaussian datasets. Then, we will provide theoretical and empirical validation for the cluster-based benchmarking approach.

\subsubsection{Generating multivariate Gaussian datasets with low-dimensional dependence structure}

The algorithm used to sample multivariate Gaussians in Figure 2 is given in \textcolor{black}{Algorithm 4}. Briefly, we generate samples of $k$ correlated dimensions by sampling $k$ bivariate Gaussians. We then complete the remaining $d-k$ ambient dimensions using half ``redundant'' dimensions (copies of the $k$ correlated dimensions) and half ``nuisance'' dimensions (independent univariate Gaussians).

\begin{algorithm}
    \caption{Generating multivariate Gaussian datasets with low-dimensional dependence structure}
    \begin{algorithmic}
        \Require ambient dimensionality $d$
        \Require dependence structure dimensionality $k$
        \Require number of nuisance dimensions $n$ (default $(d-k)/2$)
        \Require number of samples $N$
        \Require ground truth MI $b$
        \State \textbackslash\textbackslash \ Compute cov matrix to yield dependence $b$
        \State let $\rho \gets \sqrt{6*3.5} * \sqrt{1 - 2^{-2b/k}}$
        \State let $\Sigma \gets [[6, \rho], [\rho, 3/5]]$

        \State \textbackslash\textbackslash \ Sample bivariate Gaussians for  $k$ dependent dimensions
        \For{$i$ in $1..k$}
        \State let $X_i, Y_i \gets N$ samples from $\mathcal{N}(0, \Sigma)$
        \EndFor
        \State \textbackslash\textbackslash \ Duplicate random dimensions for redundant dimensions
        \For{$i$ in $1..(d-(k+n))$}
        \State let $r \gets \text{unif}([1..k])$
        \State let $X_{i+k}, Y_{i+k} \gets X_r, Y_r$
        \EndFor
        \State \textbackslash\textbackslash \ Sample univariate Gaussians for nuisance dimensions
        \For{$i$ in $0..(n-1)$}
        let $X_{d-i} \gets N$ samples from $\mathcal{N}(0, 1)$
        let $Y_{d-i} \gets N$ samples from $\mathcal{N}(0, 1)$
        \EndFor
        \State \textbf{return} [$X_1$,...$X_d$], [$Y_1$,...$Y_d$]
    \end{algorithmic}
\end{algorithm}

\subsubsection{Theoretical justification for label MI approximation of high-dimensional MI}

For the benchmarking setup described in \textcolor{black}{Section 3.2}, we have $L_y \to L_x \to X$ and $L_x \to L_y \to Y$. We will show that $I(X; Y) = I(L_x; L_y)$ under the condition that $H(L_x | X) = H(L_y | Y ) = 0$.

\begin{theorem}
    Let $L_x, L_y$ be Bernoulli random variables. Let $X, Y$ be absolutely continuous random vectors in $\mathbb{R}^N$ such that $I(L_x; Y | L_y) = I(L_y ; X | L_x) = 0$ and $H(L_x | X) = H(L_y |Y) = 0$. Then, 
    
    \begin{equation}
        I(X;Y) = I(L_x; L_y)
    \end{equation}
\end{theorem}

\begin{proof}

Due to conditional independence,

\begin{equation}
    I(L_x; L_y) = I(L_x; L_y) + I(L_x; Y | L_y)
\end{equation}

Using the chain rule for mutual information \cite{Cover2006-hn},

\begin{equation}
    I(L_x; L_y) = I(L_x; L_y) + I(L_x; Y | L_y) = I(L_x; Y) + I(L_x; L_y | Y)
\end{equation}

Applying the chain rule again,
\begin{equation}
    I(L_x; L_y) = I(L_x; Y) + I(L_x; L_y | Y) = I(X; Y) + I(L_x; Y | X) + I(L_x ; L_y | Y)
\end{equation}

Due to the non-negativity of mutual information we have

\begin{equation}
    I(L_x; L_y) \geq I(X; Y)
\end{equation}

Because $L_x, L_y$ are discrete, we can bound
\begin{equation}
    I(L_x; L_y) \leq I(X; Y) + H(L_x| X) + H(L_y | Y)
\end{equation}

So if $H(L_x | X) = H(L_y | Y) = 0$, we have

\begin{equation}
    I(L_x; L_y) \leq I(X; Y)
\end{equation}

Combining this with \textcolor{black}{(24)}, we have

\begin{equation}
    I(X; Y) = I(L_x; L_y)
\end{equation}

\end{proof}

\subsubsection{Validating the assumptions of cluster-based benchmarking setups}

The effectiveness of the benchmarking setup in \textcolor{black}{Section 3.2} relies on the assumption that $H(L_x | X) \approx H(L_y| Y) \approx 0$. We provide evidence that this is the case for MNIST 0s and 1s, and sequence embeddings of \textit{E. Coli} and \textit{A. Thaliana} proteins. First, we provide qualitative evidence by showing that label clusters (digits and species respectively) are well separated on UMAP visualizations of each dataset \textcolor{black}{(Figure 7)}. This indicates that the label of a sample can be reliably determined from its high-dimensional representation (such that $H(L_x|X) \approx 0 $). We make this notion more quantitatively precise by showing that logistic regression can predict the labels of held-out samples with high accuracy. Over 100 random $1:1$ train-test splits, logistic regression achieves mean validation accuracy $>0.98$ with standard error of mean $<10^{-3}$ for both datasets. Logistic regression classifiers are trained with $L_2$ penalty and $\lambda = 1$.

\begin{figure}
     \centering
     \begin{subfigure}[b]{0.45\textwidth}
         \centering
         \includegraphics[width=\textwidth]{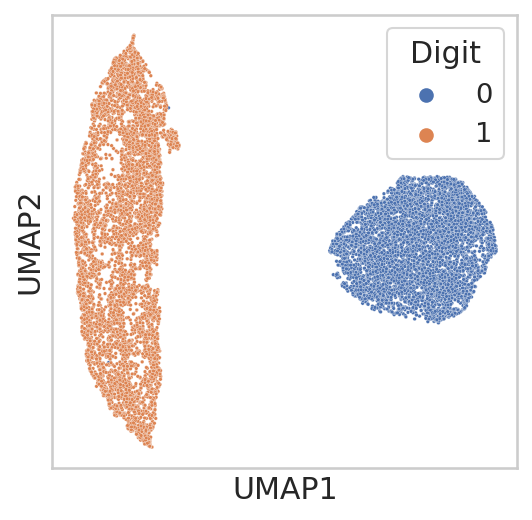}
         \caption{UMAP of binary MNIST subset.}
         \label{fig:val_mnist}
     \end{subfigure}
     \hfill
     \begin{subfigure}[b]{0.45\textwidth}
         \centering
         \includegraphics[width=\textwidth]{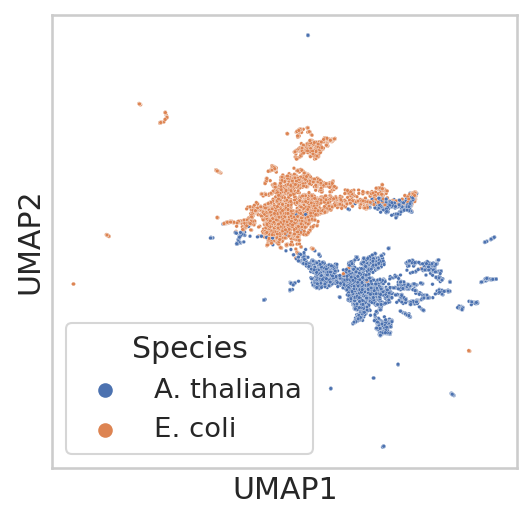}
         \caption{UMAP of proteome subsets.}
         \label{fig:val_pt5}
     \end{subfigure}
     \caption{Validating assumptions of cluster-based benchmarking}
\end{figure}

\clearpage
\subsection{Experimental reproducibility details}

Here, we will briefly summarize key details necessary for the reproduction of experiments in this paper. All of this information, and more details, can be found (albeit in a less easily readable form) in our code supplement.

\subsubsection{Preprocessing ProtTrans5 embeddings}

We downloaded ProtTrans5 embeddings of all \textit{H. sapiens}, \textit{A. thaliana}, and \textit{E. coli} proteins directly from the UniProt database. Embeddings are from \verb|prottrans_t5_xl_u50| \cite{Elnaggar2020-eq}. All proteins longer than $12\cdot 10^3$ residues are excluded. These embeddings are then unit variance normalized, and values are clipped at 10 and -10.

\subsubsection{Preprocessing hematopoiesis lineage tracing scRNA-seq data}

We first downloaded all data from Experiment 1 of the data repository from \cite{Weinreb2020-uk}. Then, we preprocessed using the Scanpy best practices \cite{Heumos2023-eh}, normalizing total reads per cell to $10^4$, log transforming, filtering for the $10^3$ most highly variable genes, and finally unit variance normalizing and clipping values at 10 and -10. We used the inferred diffusion pseudotime and SPRING embeddings computed in \cite{Weinreb2020-uk}. For pseudotime analysis, we omit cells with pseudotime value below $10^4$, because many are Lymphoid-fated rather than Neutrophil-fated. To generate joint samples of clones between two conditions, we identified all clonal barcodes that appeared in both conditions, and randomly sampled a single cell with each barcode from each condition. Because there are often several cells with the same clonal barcode in the same sample, there are many possible random pairings of clonally related cells.

\subsubsection{MINE and InfoNCE implementation details}

We use the implementations of MINE and InfoNCE from \cite{Czyz2023-qs}, with their default parameter choices. To summarize, architectures have two hidden layers with sizes (16, 8) and are optimized using Adam. 

\paragraph{Choosing critic architectures} Because the benchmark tasks of \cite{Czyz2023-qs} are dramatically different from those considered in this work (tens of dimensions as opposed to thousands), we consider the possibility that the parameters of MINE and InfoNCE used in \cite{Czyz2023-qs} are not suitable for our use case. 

To determine if other critic architectures could be more suitable, we first tested two layer architectures with layer sizes $(L, L/2)$ for various $L$ from 16 to 1024 on multivariate Gaussian data sets. We considered variables with dimensions 10, 100, and 1000, with 1 bit MI, 1-dimensional dependence structure, and $5 \cdot 10^3$ samples. We find that increasing $L$ does not improve estimation quality, and the architecture used by \cite{Czyz2023-qs} indeed is optimal for all three tested settings \textcolor{black}{(Fig. 10)}. We suspect that this is because increasing model complexity does not help when sample size remains small.

\begin{figure}
    \centering
    \includegraphics[width=0.7\textwidth]{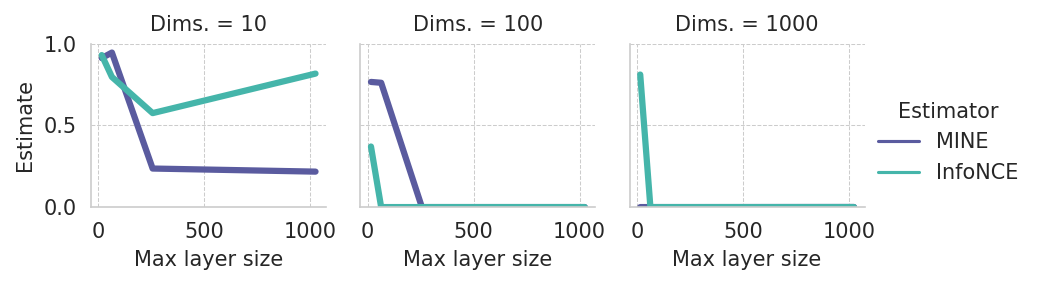}
    \caption{Variational bound estimators with increasing critic complexity, evaluated on multivariate Gaussians with 1-dimensional dependence structure. Each estimation problem has $5 \cdot 10^3$ samples and ground truth MI of 1 bit.}
    \label{fig:enter-label}
\end{figure}

On the MNIST benchmarking setup, we verify that the effectiveness of LMI estimation over MINE and InfoNCE are not merely from model complexity, by using critics with the same complexity as the LMI encoders. In line with the hypothesis that large critic architectures are not suitable for the small sample size regime, we find that with high complexity critics, MINE and InfoNCE fail entirely on the MNIST benchmark \textcolor{black}{(Fig. 11)}.

\begin{figure}
    \centering
    \includegraphics[width=0.4\textwidth]{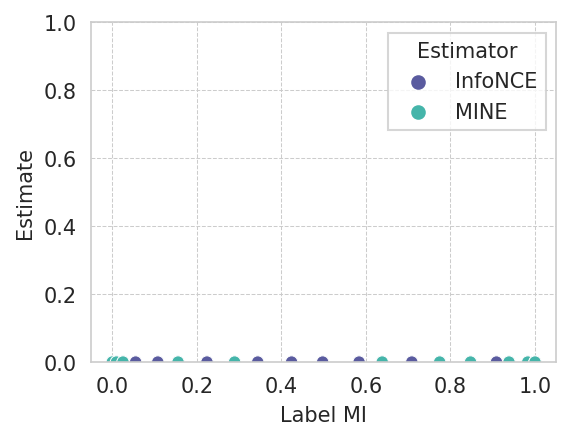}
    \caption{MNIST benchmarking for neural estimators with critic complexity equivalent to LMI encoders, over 20 datasets with true MI between 0 and 1.}
    \label{fig:enter-label}
\end{figure}

\subsubsection{KSG implementation details for pointwise decompositions}

We implement KSG with $k=3$ nearest neighbors. For protein interaction data, due to large sample numbers resulting in high computational cost, we obtain KSG estimates by averaging over estimates on batches of data containing $10^3$ samples.

We slightly adjust the KSG estimator to yield pointwise mutual information estimates. While the original KSG estimator \cite{Kraskov2004-sh} takes a sample expectation over computed pointwise mutual information values, we simply return an array of pointwise estimates rather than the average. For completeness, the algorithm is given in \textcolor{black}{Algorithm 5.}

\begin{algorithm}
    \caption{KSG estimator for pointwise estimates}
    \begin{algorithmic}
        \Require joint samples $\{(x_i, y_i)\}_{i=1}^N$
        \Require parameter $k$ (default $k=3$)
        \State let \verb|pmis| $\gets [ \hspace{1ex} ]$
        \For{each $(x_i, y_i)$}
        \State find $k$-th nearest neighbor in joint space $(x_k, y_k)$
        \State compute Chebyshev distance $d = ||(x_k, y_k) - (x_i, y_i)||_{\infty}$
        \State let $n_x \gets 0, n_y \gets 0$
        \For{each $(x_j, y_j)$}
        \If{$||x_j - x_i||_{\infty} < d$}
        \State $n_x$ += 1
        \EndIf
        \If{$||y_j - y_i||_{\infty} < d$}
        \State $n_y$ += 1
        \EndIf
        \EndFor
        \State \verb|pmis.append(| $\psi(k) + \psi(N) - \psi(n_x) - \psi(n_y)$ \verb|)|
        \EndFor
        \State \textbf{return} \verb|pmis|
    \end{algorithmic}
\end{algorithm}



\clearpage


\newpage
\section*{NeurIPS Paper Checklist}

\begin{enumerate}

\item {\bf Claims}
    \item[] Question: Do the main claims made in the abstract and introduction accurately reflect the paper's contributions and scope?
    \item[] Answer: \answerYes{} 
    \item[] Justification: The results stated in the abstract and introduction are summaries of the experimental results shown in Sections 3 and 4 (figures 2-6).

\item {\bf Limitations}
    \item[] Question: Does the paper discuss the limitations of the work performed by the authors?
    \item[] Answer: \answerYes{} 
    \item[] Justification: There is a subsection of the Discussion dedicated to limitations of the LMI approach. These limitations are considered throughout the paper. For instance, in the abstract, we highlight the necessity of low dimensional intrinsic dependence structure.

\item {\bf Theory Assumptions and Proofs}
    \item[] Question: For each theoretical result, does the paper provide the full set of assumptions and a complete (and correct) proof?
    \item[] Answer: \answerYes{} 
    \item[] Justification: Assumptions about distributions (such as absolute continuity, finite entropies) are specified in theoretical analysis (Theorems 1-4).

    \item {\bf Experimental Result Reproducibility}
    \item[] Question: Does the paper fully disclose all the information needed to reproduce the main experimental results of the paper to the extent that it affects the main claims and/or conclusions of the paper (regardless of whether the code and data are provided or not)?
    \item[] Answer: \answerYes{} 
    \item[] Justification: Key details necessary to reproduce the experimental results are given in the appendix. All experiments can be reproduced using the supplementary code. It is carefully annotated to match each result in the paper to a specific Jupyter notebook.
    

\item {\bf Open access to data and code}
    \item[] Question: Does the paper provide open access to the data and code, with sufficient instructions to faithfully reproduce the main experimental results, as described in supplemental material?
    \item[] Answer: \answerYes{} 
    \item[] Justification: As stated before, all experiments can be reproduced using the supplementary code. It is carefully annotated to match each result in the paper to a specific Jupyter notebook. The code reproducibility ``best practices'' were followed.

\item {\bf Experimental Setting/Details}
    \item[] Question: Does the paper specify all the training and test details (e.g., data splits, hyperparameters, how they were chosen, type of optimizer, etc.) necessary to understand the results?
    \item[] Answer: \answerYes{} 
    \item[] Justification: All training and test details are provided in the Appendix sections on implementation, and can be found in the supplementary code.

\item {\bf Experiment Statistical Significance}
    \item[] Question: Does the paper report error bars suitably and correctly defined or other appropriate information about the statistical significance of the experiments?
    \item[] Answer: \answerYes{} 
    \item[] Justification: We report SEMs and standard deviations in the text when relevant and feasible. No plots contain error bars.
    

\item {\bf Experiments Compute Resources}
    \item[] Question: For each experiment, does the paper provide sufficient information on the computer resources (type of compute workers, memory, time of execution) needed to reproduce the experiments?
    \item[] Answer: \answerYes{} 
    \item[] Justification: We report the hardware used for all experiments, and the projected overall environmental impact of all experiments (failed and pilot) reported in the paper. We do not provide further granularity because the experiments have rather modest compute requirements -- the entirety of the experiments in this paper can be reproduced using a commercial NVIDIA GPU (RTX 3090) in about one day (by running every Jupyter notebook in the code supplement).
    
    
\item {\bf Code Of Ethics}
    \item[] Question: Does the research conducted in the paper conform, in every respect, with the NeurIPS Code of Ethics \url{https://neurips.cc/public/EthicsGuidelines}?
    \item[] Answer: \answerYes{} 
    \item[] Justification: We have reviewed and adhered to the guidelines. We mention potential social and environmental impacts in the Discussion section.

\item {\bf Broader Impacts}
    \item[] Question: Does the paper discuss both potential positive societal impacts and negative societal impacts of the work performed?
    \item[] Answer: \answerYes{} 
    \item[] Justification: We discuss the potential social impact of MI estimators.
    
\item {\bf Safeguards}
    \item[] Question: Does the paper describe safeguards that have been put in place for responsible release of data or models that have a high risk for misuse (e.g., pretrained language models, image generators, or scraped datasets)?
    \item[] Answer: \answerNA{} 
    \item[] Justification: We do not release data or models with high risk for misuse.

\item {\bf Licenses for existing assets}
    \item[] Question: Are the creators or original owners of assets (e.g., code, data, models), used in the paper, properly credited and are the license and terms of use explicitly mentioned and properly respected?
    \item[] Answer: \answerYes{} 
    \item[] Justification: We cite the library used to implement the LMI approximation (PyTorch).

\item {\bf New Assets}
    \item[] Question: Are new assets introduced in the paper well documented and is the documentation provided alongside the assets?
    \item[] Answer: \answerYes{} 
    \item[] Justification: The code developed for latent MI approximation (the \verb|lmi| library provided in the supplement) is well documented.

\item {\bf Crowdsourcing and Research with Human Subjects}
    \item[] Question: For crowdsourcing experiments and research with human subjects, does the paper include the full text of instructions given to participants and screenshots, if applicable, as well as details about compensation (if any)? 
    \item[] Answer: \answerNA{} 
    \item[] Justification: This work does not involve human subjects or crowdsourcing.

\item {\bf Institutional Review Board (IRB) Approvals or Equivalent for Research with Human Subjects}
    \item[] Question: Does the paper describe potential risks incurred by study participants, whether such risks were disclosed to the subjects, and whether Institutional Review Board (IRB) approvals (or an equivalent approval/review based on the requirements of your country or institution) were obtained?
    \item[] Answer: \answerNA{} 
    \item[] Justification: This work does not involve human subjects or crowdsourcing.

\end{enumerate}

\end{document}